\documentclass[prl,showpacs, twocolumn, preprintnumbers]{revtex4}%

\usepackage{amssymb,hyperref, amsmath}
\usepackage[dvips]{color}
\usepackage[dvips]{graphicx}
\usepackage{epsfig}

\begin{document}

\preprint{JLAB-THY-06-512}

\title{Two Photon Decays of Charmonia from Lattice QCD}

\author{Jozef J. Dudek}
\affiliation{Jefferson Laboratory MS 12H2, 12000 Jefferson Avenue, Newport News, VA 23606, USA}
\email{dudek@jlab.org}

\author{Robert G. Edwards}
\affiliation{Jefferson Laboratory MS 12H2, 12000 Jefferson Avenue, Newport News, VA 23606, USA}
\email{edwards@jlab.org}

\begin{abstract}
We make the first calculation in lattice QCD of two-photon decays of
mesons. Working in the charmonium sector, using the LSZ
reduction to relate a photon to a sum of hadronic vector eigenstates, 
we compute form-factors in both the space-like and time-like domains for the transitions $\eta_c \to \gamma^*
\gamma^*$ and $\chi_{c0} \to \gamma^* \gamma^*$. At the on-shell point
we find approximate agreement with experimental world-average values.
\end{abstract}

\pacs{12.38.Gc, 12.40.Vv, 13.25.Gv, 14.70.Bh}

\maketitle 


The two photon widths of charmonia have attracted the attention of a
range of a theoretical and experimental techniques. Within
perturbative QCD the two-photon branching fraction is
believed to give access to the strong coupling constant at the
charmonium scale though cancellation of non-perturbative factors, while in quark models and the effective field theory NRQCD, two-photon widths have been proposed as a
sensitive test of the corrections to the non-relativistic
approximation. Experimental measurements are diverse, coming from on
the one hand two-photon fusion at $e^+ e^-$ machines with subsequent
reconstruction of the charmonium in light hadrons, and, on the other,
$p\bar{p}$ annihilation to charmonium with decay to real
$\gamma \gamma$ pairs being detected. Improvements in
extracted branching fractions from both methods will soon arrive via
more accurate measurements from Belle, BaBar and CLEO-c of exclusive hadronic $\eta_c$ decays.

Despite the considerable theoretical and experimental interest, there
has not yet been an ab-initio estimation of two photon widths in charmonia.
In this paper we address this
issue using a novel application of lattice QCD.

Conventionally lattice QCD calculations involve the evaluation of the matrix element of
an operator between hadron states. The hadrons are induced using
interpolating fields (some combination of quark and gluon fields) with
the appropriate hadron quantum numbers. However these fields will typically
have overlap with many excited states which have the same external quantum
numbers so that in order to isolate the ground states, the
interpolating fields are taken out to Euclidean times far from the
operator (and each other). Such an approach will clearly not work if
the initial or final state contains no hadron as is the case for two-photon
decays since the photon is not an eigenstate of QCD. In this case we can adopt a slightly more sophisticated
approach, using the formal relationship between the $S$-matrix and
field-theoretic $N$-point functions and the accurate perturbative expansion in the
photon-quark coupling to express the photon as a superposition of QCD eigenstates.

The concept and realization is lucidly presented by Ji and Jung in
\cite{Ji:2001wh, Ji:2001nf} where they
consider the hadronic structure of the photon using lattice
three-point functions. We will outline the method as applied to meson
two-photon decays.

The amplitude for the two-photon decay of a meson $M$ can be expressed
in terms of a photon two-point function in Minkoswki space by means of
the LSZ reduction
\begin{multline}
 \langle \gamma(q_1, \lambda_1)\gamma(q_2, \lambda_2) | M(p) \rangle
 = - \lim_{\substack{q'_1 \to q_1 \\ q'_2 \to q_2}} \epsilon^*_\mu(q_1,
 \lambda_1) \epsilon^*_\nu(q_2, \lambda_2)\nonumber \\
\times q_1'^2 q_2'^2 \int d^4x d^4y \, e^{i q'_1.y +
 i q'_2.x} \langle 0 | T\big\{ A^\mu(y) A^\nu(x) \big\} | M(p) \rangle,\nonumber
\end{multline}
up to photon renormalisation factors. The explicit photon fields
prevent direct computation of this quantity in lattice QCD, however we
can utilize the perturbative expansion of the photon-quark coupling to
approximately integrate them out (the path-integral over gluon fields
is suppressed): $\int {\cal D}A {\cal D} \bar{\psi} {\cal D}\psi e^{i S_{QED}[A, \bar{\psi}, \psi]} A^\mu(y) A^\nu(x) =$
\begin{widetext}
\begin{equation}
 \int {\cal D}A {\cal D} \bar{\psi} {\cal D}\psi  e^{i S_0[A, \bar{\psi}, \psi]}
 \big( \ldots + \tfrac{e^2}{2}\int d^4z d^4w \;
 \left[\bar{\psi}(z)\gamma^\rho \psi(z) A_\rho(z)\right] \; \left[\bar{\psi}(w)\gamma^\sigma \psi(w) A_\sigma(w)\right]  +\ldots \big)   A^\mu(y) A^\nu(x), \nonumber
\end{equation} 
The integration over the photon field can be carried out by Wick
contracting the fields into propagator products, so that, neglecting
disconected pieces, $\langle \gamma(q_1, \lambda_1)\gamma(q_2, \lambda_2) | M(p) \rangle =$
\begin{multline}
 {(-e^2)} \lim_{\substack{q'_1 \to q_1 \\ q'_2 \to q_2}} \epsilon^*_\mu(q_1,
 \lambda_1) \epsilon^*_\nu(q_2, \lambda_2) q_1'^2 q_2'^2
\int d^4x\, d^4w\, d^4z \, e^{i q'_1.x}\, D^{\mu \rho}(0,z) D^{\nu \sigma}(x, w) \langle
 0 | T\big\{ j_\rho(z) j_\sigma(w)\big\} | M(p)\rangle.\nonumber
\end{multline}

The photon propagator can be written $D^{\mu\nu}(0,z) = -i
g^{\mu\nu} \int \tfrac{d^4k}{(2\pi)^4}\tfrac{e^{i k.z}}{k^2 + i
  \epsilon} $, cancelling the inverse propagators outside the integral.

\end{widetext}
As explained in \cite{Ji:2001wh}, the resulting expression can be analytically
continued from Minkowski to Euclidean space-time provided the photon
virtualities, $Q_1^2 = |\vec{q}_1|^2 - \omega_1^2$, $Q_2^2 =
|\vec{q}_2|^2 - \omega_2^2$ are not sufficiently timelike that they
can produce on-shell hadrons. In charmonium\footnote{This is true within the quenched truncation,
  neglecting disconected diagrams. Relaxing these approximations
  allows production of light vector mesons or vector glueballs -
  phenomenologically we expect these states to have small coupling to
  the charmonium meson} this limits us to $Q^2 > -
m_{J/\psi}^2$. Using suitable a QCD interpolating field to produce $M$ and
reversing the operator time-ordering for convenience we have $\langle M(p) | \gamma(q_1, \lambda_1) \gamma(q_2, \lambda_2)\rangle =$
\begin{widetext}
\begin{equation}
\lim_{t_f-t \to \infty} e^2 \frac{ \epsilon_\mu(q_1,
 \lambda_1) \epsilon_\nu(q_2, \lambda_2)}{\tfrac{Z_M(p)}{2 E_M(p)} e^{-E_M(p)(t_f-t)}} \int dt_i
e^{- \omega_1 (t_i -t)}
\langle 0 | T\Big\{ \int d^3 \vec{x}\, e^{-i\vec{p}.\vec{x}}
\varphi_M(\vec{x}, t_f) \int d^3 \vec{y}\, e^{i\vec{q_2}.\vec{y}}
j^\nu(\vec{y}, t) j^\mu(\vec{0}, t_i)       \Big\}|0\rangle  \label{master}
\end{equation}
\end{widetext}

It is clear from the previous discussion that obtaining two-photon
widths is a natural extension to the study of radiative transitions
carried out in \cite{Dudek:2006ej} - there we computed three-point
functions involving vector currents with the source $(t_i)$ and sink $(t_f)$ positions
fixed and varied the vector current ``insertion'' ($t$) across the temporal
direction to plot out a plateau. For two-photon widths we repeat this but with a varying sink (or in the case of
eqn. \eqref{master}, source) position which will be
integrated over with an exponential weighting. 

Our calculation was performed in the quenched truncation of QCD using
the Chroma software system. We employed a $24^3 \times 48$ lattice
generated using a Wilson gauge action with coupling $\beta=6.5$ and
lattice spacing $a \approx 0.047$ fm (determined from the static quark
potential in \cite{Edwards:1997xf}). 
Quark propagators were computed
using a non-perturbatively improved Clover
action~\cite{Luscher:1996ug} with Dirichlet boundaries in the temporal
direction, with the charm quark mass set using the
  spin-averaged $\eta_c, J/\psi$ mass, accurate to within 3\%. Since we adopted only smeared local fermion bilinears as
meson interpolating fields we had access to three-point functions
involving $\eta_c$ and $\chi_{c0}$ mesons. In \cite{Dudek:2006ej}
typical radii of charmonium states were extracted and found to be much
smaller than the $\sim 1.2$ fm box used here. We adopt 
the conserved (``point-split'') vector current at the insertion - our
experience in radiative transitions is that including the improvement
term\cite{Martinelli:1990ny} has an effect below 5\% near $Q^2 =0$.

\begin{figure}[h]
 \centering
\vspace{-9mm}
\psfig{width=8.5cm,file=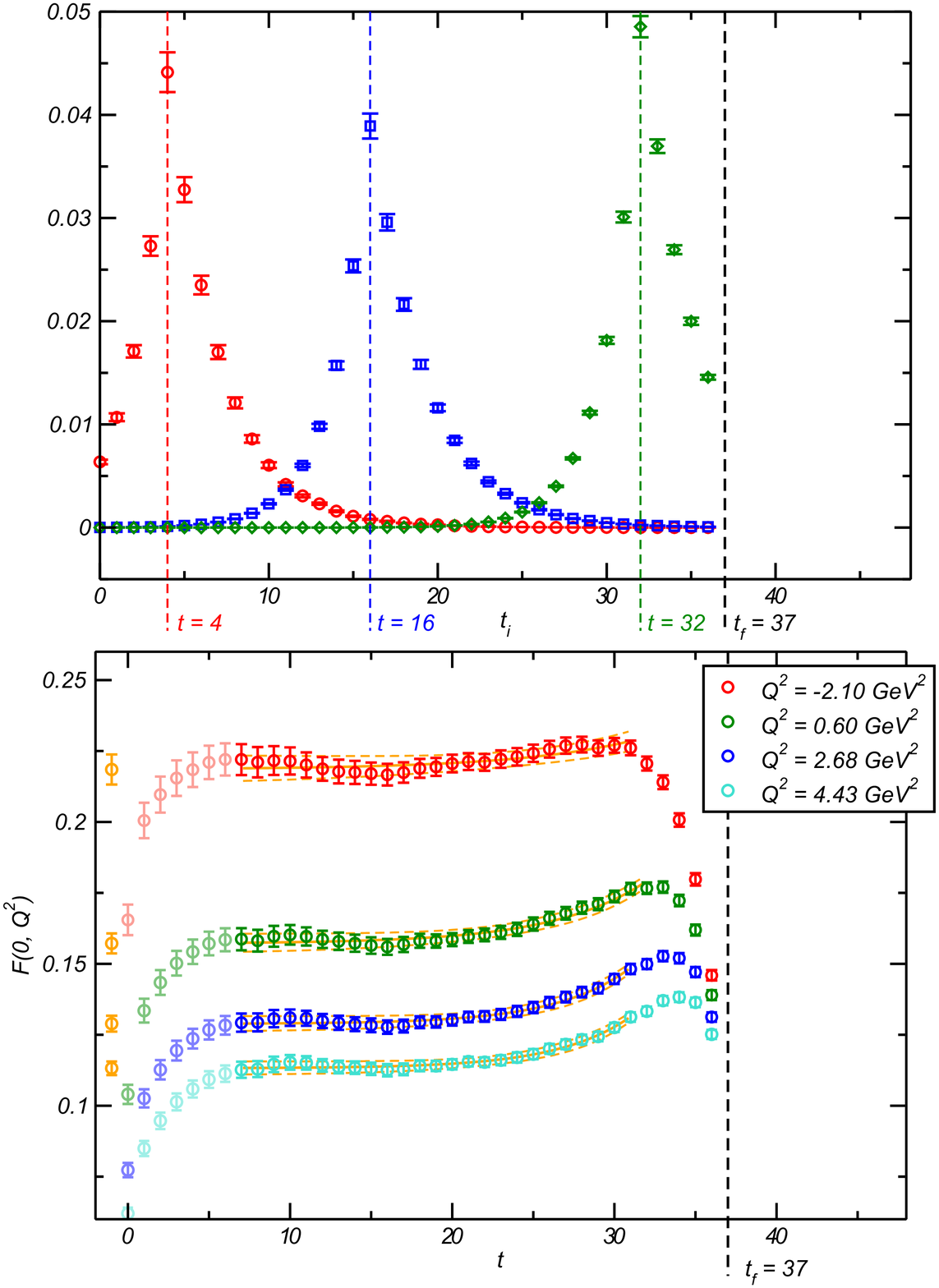}
 \caption{(a) Integrand in \protect{ \eqref{master} } at three values
   of vector current insertion time ($t=4,16,32$) with pseudoscalar
   sequential source at sink position $t_f=37$. (b) Pseudoscalar
   two-photon form-factor as a function of time slice, $t$, from
   \protect{ \eqref{master}}. First six time slices ghosted out due to
   the Dirichlet wall truncating the integral. Constant plus single
   exponential fits shown in orange.}
 \label{fig:integrand}
\end{figure}

We applied two different methods to calculate \eqref{master}. The
first (using 174 configurations) was to place the meson state at a
fixed sink position $t_f=37$. As in \cite{Dudek:2006ej} the sink was
used as a sequential source for a backward propagator inversion, meaning that
its properties were fixed for each computation while we were able, without
further cost, to vary the direction and momentum of the insertion and
the direction of the source field. We then computed with all possible
source positions, $t_i$, which, while costly in computing time, allowed us to
freely vary the value of $\omega_1$ and hence $Q^2_1$ and in addition to view the
subsequent integrand. In figure \ref{fig:integrand}(a) we display the integrand for
``insertion'' positions $t=4,16,32$, $\vec{p}_f=(000)$ and $\vec{q_1}=(100)$
with an $\eta_c$ at the sink.

\begin{figure}[h]
 \centering
\psfig{width=8.7cm,file=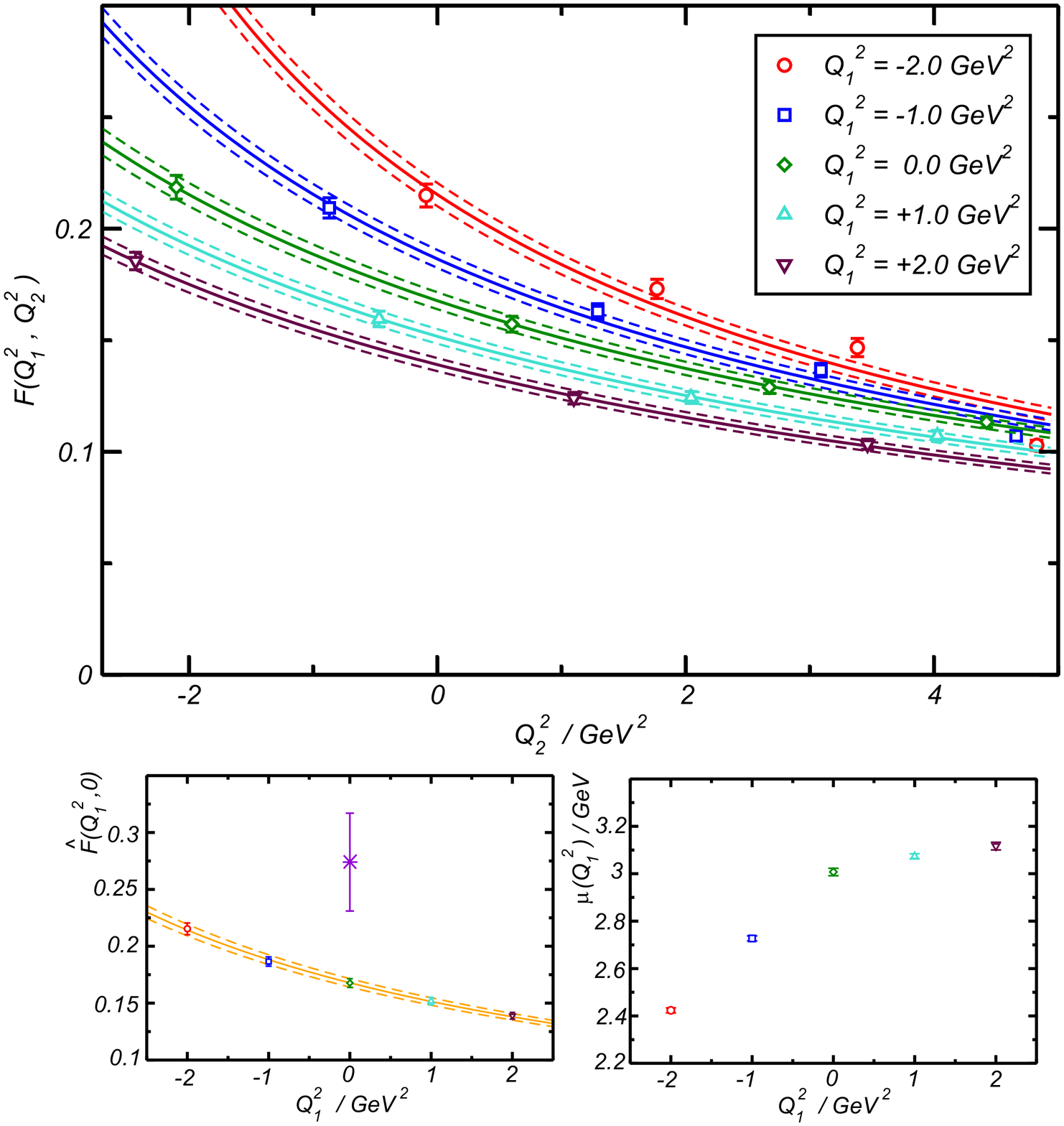}        
 \caption{$\eta_c$ two-photon form-factor, $F(Q_1^2, Q_2^2)$.  Points
   are lattice QCD data, fits are monopole forms as described in the
   text. 
Lower left shows fitted amplitudes at
   $Q_2^2=0$ and lower right the fitted pole masses $\mu(Q_1^2)$. }
 \label{fig:F}
\end{figure}

It is clear that provided the insertion is not placed too close to the
Dirichlet wall (i.e. $t \gtrsim 7$) we will be able to capture the full
integral by summing time slices, $t_i$. The integral as a function of
insertion position $t$ is shown in figure \ref{fig:integrand}(b) for a selection of
$Q_2^2$ with $Q_1^2=0$ ($\omega_1 = |\vec{q}_1|$) where we observe plateaus with the deviation from
plateau behavior at larger $t$ coming from excited $\eta_c$
contributions, both of which are fitted simultaneously. Extracting the plateau
values for a range of $Q_1^2$ (which we are free to choose
continuously) and $Q_2^2$ (which is fixed for a given set of $\omega_1,
\vec{q}_2, \vec{p}$), we find the dependence displayed in figure
\ref{fig:F}. We plot dimensionless $F$ defined by $\langle \eta_c| \gamma(q_1,
\lambda_1)\gamma(q_2, \lambda_2)\rangle = 2 (\tfrac{2}{3} e)^2 m_{\eta_c}^{-1} F(Q_1^2,
  Q_2^2)  \epsilon_{\mu\nu\rho\sigma} \epsilon^\mu_1
\epsilon^\nu_2 q_1^\rho q_2^\sigma$, where the on-shell decay width is
$\Gamma(\eta_c \to \gamma \gamma) = \pi \alpha_{em}^2 \tfrac{16}{81}
m_{\eta_c} |F(0,0)|^2$. 

In order to more clearly express the voluminous data in figure \ref{fig:F} we adopt a
simple one-pole parameterization to fit the data for each value of $Q_1^2$,
\begin{equation}
F(Q_1^2, Q_2^2) = \frac{\hat{F}(Q_1^2, 0)}{1+ \tfrac{Q_2^2}{\mu^2(Q_1^2)}}.
\end{equation}
The curves in figure \ref{fig:F} are fits of this form. In the insets
of figure \ref{fig:F} we display the fit parameters $\hat{F}(Q_1^2, 0), \mu(Q_1^2)$. $\hat{F}(Q_1^2, 0)$
is itself then fitted with a one-pole form with $\hat{F}(0,0) = 0.168(4),
\hat{\mu} = 3.040(20)\, \mathrm{GeV}$, which is compatible with
$\mu(0) = 3.008(16)\, \mathrm{GeV}$ indicating the reasonableness of
the one-pole parameterization in this $Q^2$ region. The both-photons
on-shell point is three standard deviations smaller than the value
extracted from the PDG\cite{Eidelman:2004wy} decay width $=0.274(43)$. 

The error quoted on our result is statistical only and must be
augmented by an uncertainty due to scaling from our fixed lattice
spacing to the continuum and one related to the lack of light-quark
loops within the quenched truncation. Since we compute at only one
lattice spacing we cannot accurately determine a scaling uncertainty,
however we expect the non-perturbatively tuned Clover action to have
small ${\cal O}(a)$ scaling errors. On the other hand it is possible that we
have non-negligible ${\cal O}(m_c a)$ scaling which we conservatively estimate to be at the
15\% level.

In the NRQCD factorized approach the decay amplitude is proportional
to the $\eta_c$ wavefunction at the origin with relativistic
corrections. In the quenched theory utilized here, the incorrect running of the
coupling down to short distances with the scale set at long distances,
does typically lead to a depletion of the wavefunction at the
origin. On the lattice used here we find\footnote{The $\eta_c$
  decay constant should be equal to the $J/\psi$ up to small
  spin-dependent corrections, see \cite{Dudek:2006ej} for an explicit lattice extraction.} $f_\psi = 386(6)$ MeV in comparison to
the experimental 411(7) MeV, however we cannot determine to what
degree the quenched depletion is offset by ${\cal O}(m_c a)$ scaling effects,
and as such we assign an estimated 10\% error due to quenching.  

The similarity of $\mu(0)$ to the $J/\psi$ mass in our calculation
($3006(5)$ MeV) suggests an approximate vector meson dominance description of the
form-factor. This is at first sight somewhat surprising given that VMD
was demonstrated not to hold for the $\eta_c$ charge ``form-factor''
in \cite{Dudek:2006ej}. This was explained as being due to the presence of
very many closely spaced vector poles (the $J/\psi, \psi' \ldots$)
which must be summed -  in contrast to the light-quark sector where the
$\rho(1450)$ is rather distant in comparison to the $\rho(770)$ and
where VMD tends to be a good approximation. 

What we suspect is happening here is that while the excited $\psi$
poles are not negligible by their remoteness, they are almost negligible by
their small residues. Consider the case $Q_1^2=0$ and the time
ordering in which photon 1 is emitted from the $\eta_c$ first, then
the time-dependent perturbation theory amplitude is proportional to 
\begin{equation}
\sum_N  \frac{  \langle \eta_c | j_\mu(0) |\psi^{(N)}\rangle \langle
  \psi^{(N)} | j_\nu(0) | 0\rangle}{m(\eta_c) - E(\psi^{(N)})}.
\end{equation}
The numerator is essentially the residue of the $N$th excited $\psi$
pole and is seen to be proportional to the product of the $M1$ transition matrix
element between $\eta_c$ and $\psi^{(N)}$ and the decay constant of
the $\psi^{(N)}$. The decay constants fall slowly with
increasing $N$ as observed from the experimental $e^+e^-$ widths,
while the $M1$ transition amplitudes are expected to fall rather
rapidly - they are ``hindered'' transitions that are proportional to
the overlap of orthogonal wavefunctions with a small correction for
recoil. Because of this only the $J/\psi$ term in the sum has a
considerable residue (since the wavefunction overlap with the $\eta_c$
is close to one) and we observe something like VMD. A hint of this
behavior was observed in the L3 experiment \cite{Acciarri:1999rv}, and a model
realization demonstrated in \cite{Lakhina:2006vg}.

The fact that $\mu$ varies with $Q_1^2$ suggests that as this photon
goes off-shell the higher $\psi$ resonances get a larger residue
(through a larger recoil correction to the hindered transition) and
contribute more to the sum over poles moving the effective pole
position, $\mu(Q_1^2)$.  

\begin{figure}[h]
 \centering
\vspace{-0.7cm}
\psfig{width=8.5cm,file=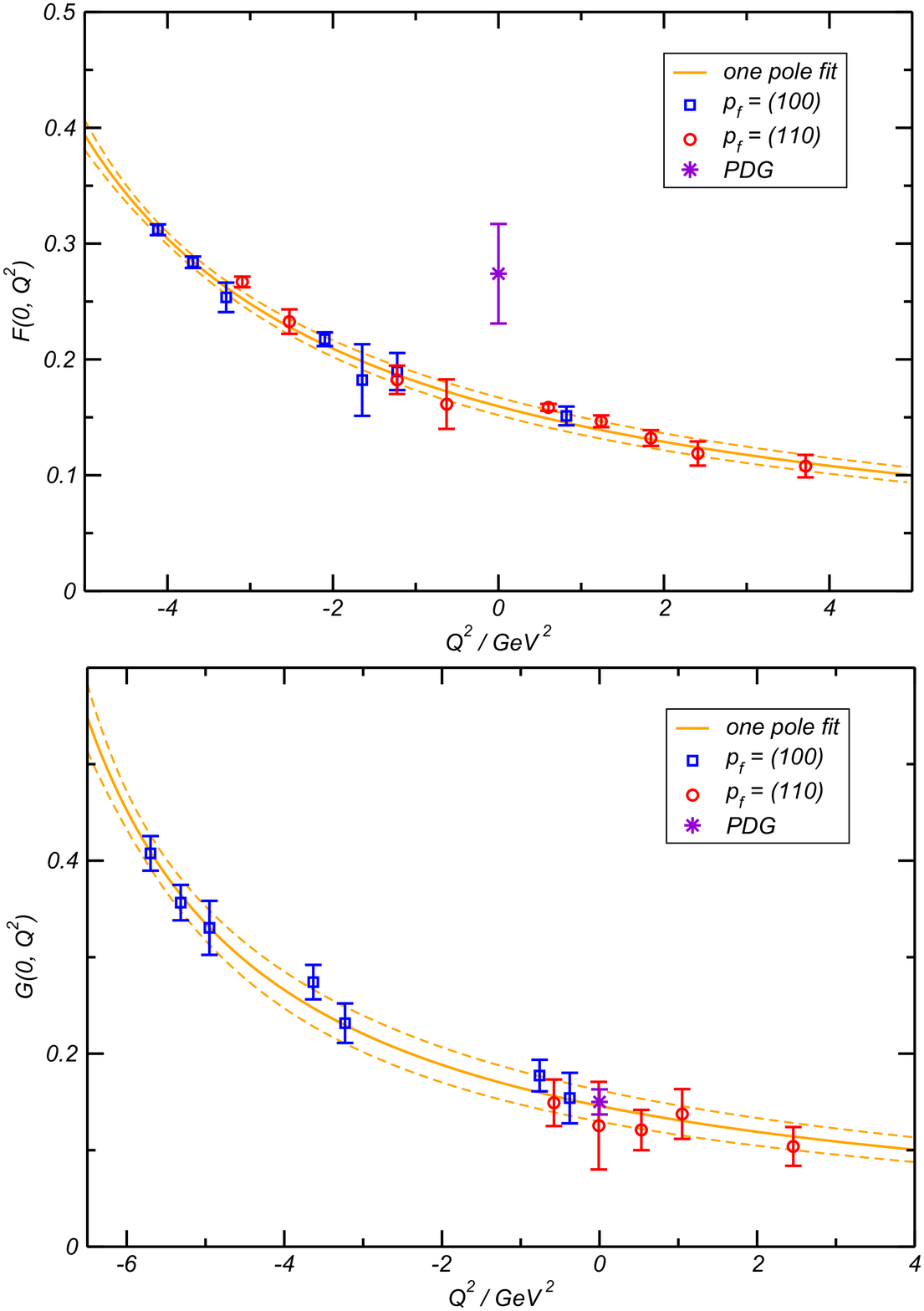}
 \caption{(a) $\eta_c \to \gamma \gamma^*$ amplitude. (b) $\chi_{c0}
   \to \gamma \gamma^*$ amplitude. Fits are monopole forms as described in the text.}
 \label{fig:fast}
\end{figure}

The method of computing three-point functions with all possible
source positions ($t_i = 0 \ldots t_f$) is extremely costly in computing resources. With the
penalty of losing the ability to freely vary $Q_1^2$ we can reduce the computing
time by a factor of ${\cal O}(L_t)$ by putting the meson interpolating
field at the source and using 
\begin{equation}
\int dt e^{\omega_1 t} \int d^3\vec{z}\, e^{i\vec{q}_1.\vec{z}}
\bar{\psi}(\vec{z}, t) \gamma^\mu \psi(\vec{z}, t)
\end{equation}
in the sequential source for the backward propagator inversion. It is
then necessary to fix $\omega_1$ and $\vec{q}_1$ in advance and one cannot
view the integrand since the integration is being performed
``on-the-fly'' within the sequential source. We computed $300$ configurations with
$Q_1^2=0$ in this way and the results for $\eta_c$ and $\chi_{c0}$
are displayed in figure \ref{fig:fast}. Note that the one-pole fit to
the $\eta_c$ data yields $F(0,0) = 0.160(8)$ and $\mu = 2.899(71)$, values that are in agreement with
those extracted from the method described above.

We define the $\chi_{c0}$ two-photon form-factor by 
\begin{multline}\langle
\chi_{c0}|\gamma(q_1, \lambda_1) \gamma(q_2, \lambda_2) \rangle = 2
(\tfrac{2}{3} e)^2 \\ \times m_{\chi_{c0}}^{-1} G(Q_1^2, Q_2^2) \big(
\epsilon_1 \cdot \epsilon_2 q_1 \cdot q_2 - \epsilon_2 \cdot q_1
\epsilon_1 \cdot q_1 \big).
\end{multline}
The PDG\cite{Eidelman:2004wy} average of experimental
measurements for the two-photon width corresponds to $|G(0,0)| =
0.150(13)$ with which we are in excellent agreement - our single-pole
fit to lattice data returns $|G(0,0)| = 0.146(16)$ and $\mu =
2.976(81)$ GeV.  

The logic used to justify the approximate $J/\psi$ VMD observed for the $\eta_c$
does not so obviously apply to the $\chi_{c0}$. The radiative
transitions to virtual vector meson states are now
$E1$ and are not ``hindered'' for the excited states to the same
extent as $M1$ (This can be seen in, for example, the quark model
study \cite{Barnes:2005pb}). Since
the residues of excited state poles are not falling so quickly they
can contribute to the sum and we will not necessarily see dominance of the $J/\psi$.

Unfortunately in this case the two nearest poles, the $\psi(3686)$ and the $\psi(3770)$ can
conspire to cancel each other's effect. Using a combination of
experimental data and quark model predictions\cite{Barnes:2005pb} one estimates that the
residues of these two poles are approximately equal (they have value of
about $1/3$ the $J/\psi$ residue). If these residues happen to have opposite
sign then the $J/\psi$ pole will dominate with small contributions from
the more highly excited $\psi$ resonances. Our data is insufficiently
precise to distinguish any deviation from a one-pole behavior. 


In summary we have demonstated the feasibility of using a suitable
sum over lattice timeslices to simulate a photon in an external state,
using the phenomenologically interesting case of the two-photon decays
of charmonia. Our results 
$\Gamma( \eta_c \to \gamma \gamma ) =
2.65(26)_{\mathrm{stat}}(80)_{\substack{\mathrm{scal.} \\ \mathrm{sys}}}(53)_{\substack{\mathrm{quen.} \\ \mathrm{sys}}}\, \mathrm{keV}$, $\Gamma(\chi_{c0} \to \gamma
\gamma) = 2.41(58)_{\mathrm{stat}}(72)_{\substack{\mathrm{scal.} \\
    \mathrm{sys}}}(48)_{\substack{\mathrm{quen.} \\ \mathrm{sys}}}\,\mathrm{keV}$, are in reasonable agreement with experiment. We believe them to be systematics dominated, in
particular owing to the neglect of light quark loops and
discretisation effects related to the heavy quark mass. The
systematic error can be reduced with further computations, which are
now warranted given the feasibility demonstrated herein.
This lattice technique has the advantage over non-relativistic models that it
can be applied to the light-quark sector without fundamental change.

\begin{acknowledgments}


Notice: Authored by The Southeastern Universities Research
Association, Inc. under U.S. DOE Contract No. DE-AC05-84150. The
U.S. Government retains a non-exclusive, paid-up, irrevocable,
world-wide license to publish or reproduce this manuscript for
U.S. Government purposes.  
Computations were performed on clusters at Jefferson Laboratory
under the SciDAC initiative.

\end{acknowledgments}


\end{document}